\documentclass[11pt, twocolumn]{IEEEtran}
\usepackage{amsmath, amssymb, color, graphicx, float, multirow, adjustbox, caption}
\usepackage{algorithm}
\usepackage[noend]{algpseudocode}
\usepackage{courier}
\usepackage{authblk}

\usepackage{hyperref}
\hypersetup{colorlinks=true, urlcolor=blue}

\newcommand\numberthis{\addtocounter{equation}{1}\tag{\theequation}}

\title{User-Adaptive Text Entry for Augmentative and Alternative Communication}
\author{Matt Higger\thanks{Our work is supported by NSF (IIS-1149570, CNS-1544895), NIDLRR (90RE5017-02-01), and NIH (R01DC009834).}}
\author{Fernando Quivira}
\author{Deniz Erdogmus}
\affil{Electrical and Computer Engineering, Northeastern University}

\begin{document}
\maketitle

\begin{abstract}
The viability of an Augmentative and Alternative Communication device often depends on its ability to adapt to an individual user's unique abilities.

Though human input can be noisy, there is often structure to our errors.  For example, keyboard keys adjacent to a target may be more likely to be pressed in error.  Furthermore, there can be structure in the input message itself (e.g. `u' is likely to follow `q').  In a previous work, `Recursive Bayesian Coding for BCIs' (IEEE Transactions on Neural Systems and Rehabilitation Engineering, 2016), a query strategy considers these structures to offer an error-adaptive single-character text entry scheme.  However, constraining ourselves to single-character entry limits performance.  A single user input may be able to resolve more uncertainty than the next character has.  In this work, we extend the previous framework to incorporate multi-character querying similar to word completion.  During simulated spelling, our method requires $20\%$ fewer queries compared to single-character querying with no accuracy penalty.  Most significantly, we show that this multi-character querying scheme converges to the information theoretic capacity of the discrete, memoryless user input model.
\end{abstract}


\section{Introduction}\label{sec:intro}
Many people do not have the physiology to express themselves by traditional means, Brain Computer Interfaces (BCIs) offer a voice built from the physiology a user can produce.  For example, some BCIs empower the user to type a message by producing a sequence of imagined movements measured via Electroencephalography (\cite{Blankertz2006, Akcakaya2014}).   A central question of BCIs is how to provide efficient communication access to people whose abilities vary from slightly impaired dexterity to total paralysis.

Many of these problems can be posed as discrete inference.  From the $N$ physiological responses a user can produce they generate a sequence associated with the message they want to communicate.  While response capability and consistency may vary, reliabile autonomous communication is a near universal goal \cite{Hawley2008, Mobasheri2016}.

Switch Scanning is a popular approach as it requires only that a user can reliably produce a single, timed input.  A system suggests sets of letters and the user creates a response when their target is present.  This method has been optimized to minimize the expected number of queries per letter decision \cite{higger_2016_karp} as well as having timing and layout factors tuned to optimize typical Human Computer Interface (HCI) concerns \cite{Francis2011}.  Users who are capable can improve performance by using a system which accepts more than just one response such as a decision tree which accepts eye gaze gestures \cite{Cecotti2016} or a Huffman decision tree which accepts finger flexions \cite{Bajer2012}.

Some performance boost can be had for `free' (without burdening the user or modifying the classification scheme) by incorporating a language model \cite{Mora-Cortes2014}.  Many systems (\cite{Felzer2013, Felzer2015, Volosyak2011}) dynamically switch into a word completion mode after observing the first few typed characters.  Task switching from character to word selection often increases the attention demands of the user, Quinn et al. take care to ensure the word completion benefit justifies its cost \cite{Quinn2016}.

Ambiguous keyboard schemes avoid some of these costs by moving on to query the next letter even if the user's input leaves some ambiguity about the current letter \cite{Judge2011} (e.g. T9, \cite{king2001reduced}).  Combined with switch scanning systems, ambiguous style keyboards show significant promise \cite{K2003, Felzer2010}.  

Of course, errors may happen and many systems, including those mentioned above, incorporate a backspace to undo previous decisions.  Fowler et al. consider the full history of user errors to provide an intelligent backspace that is capable of removing more than just the most recently decided character \cite{Fowler2013}.

Cental to many BCI typing systems is the challenge of query pacing.  Should we ask the user about the current, uncertain part of their message or move on to future letters and disambiguate later?  To mitigate how often the pacing problem arises Ahani et al. query on the level of the word rather than letter \cite{Ahani2014}.  Alternatively, it is possible to remove or rearrange words within a message while still retaining its meaning \cite{Jaeger2016}; we might do well to choose the best paced target message which still preserves the user's meaning.  In this work, we mitigate the pacing problem by querying the user about a variable number of letters beyond their current candidate message.

The rate of Mutual Information between the user's intended and estimated inputs, Information Transfer Rate (ITR), is a widely used (\cite{Dornhege}) performance metric.  When computed correctly \cite{Yuan2013, Roberts2000, Nykopp2001, Fatourechi2006, Dornhege} ITR describes the rate of uncertainty reduction in a target variable.

It is instructive to consider a typing system which fails to meet its capacity.  Let us assume English has an entropy of 2 Bits per character\footnote{This is not far off, English is estimated to have at most 1.75 Bits per character \cite{Brown1992}} and a user who has an errorless binary input channel.  Technically, this user is capable of averaging a character with each two inputs although the communication scheme that achieves this performance is not obvious.  Alphabetically querying (``Is your target letter A?'', ``Is your target letter B?'', ...) is anticipated to average more than two questions to identify a letter  (`A' and `B' are the only letters which take two or fewer questions).  These alphabetical queries struggle as they are often answered with an unsurprising `no, not my character yet.'  Information is only derived from questions whose responses are surprising.   In other words, a scheme's inference rate is limited by the a priori uncertainty in the evidence it will receive.  One cannot learn what is already known.   

Lesher \cite{Lesher1998} optimizes for errorless, discrete (not necessarily binary) inputs.  Walmsley \cite{Walmsley2014} offers a static (not changing to adapt to the local context of the user's typed message) solution for discrete inputs with errors.  Further work \cite{higger_2016_tnsre, Akce205} incorporates a user error model and the local context of a user's typed message to construct better queries.  Characters are associated with user inputs to minimize the expected uncertainty remaining in the target variable after incorporating the user's response.  While an improvement, the single-character constraint precludes this system from achieving the capacity of the user's input channel.  To see this, consider a system which decides on the $n$th target character when that letter's entropy falls below some threshold.  It is possible this next letter is vague enough to continue querying but not so uncertain as to satisfy the user's input capacity.  In a work most similar to our own, Omar et al address multi-letter querying for binary inputs \cite{Omar2011}.

In this work, we relax the single-letter constraint to approach the user's $n$-ary input capacity.  Intuitively, we satisfy the user's capacity by adding uncertainty from future characters when the next character is almost known.  We formalize this intuition (Sec \ref{sec:method}), describe an example instantiation (Sec \ref{sec:web_speller}), simulate the performance benefit (Sec \ref{sec:simulation}), and show that this new scheme is capable of achieving the user's capacity (Sec \ref{sec:capacity}).  Please note that our contribution is identifying a querying and pacing scheme which is optimal from an inference standpoint.  While Human Computer Interface considerations (see \cite{MacKenzie2013, Quinn2016}) should be considered before a final system is designed, they are not the focus of this work.

\section{Method}\label{sec:method}
If possible, we suggest viewing a video\footnote{\href{https://www.youtube.com/watch?v=npsL-upO5Ww}{\underline{https://www.youtube.com/watch?v=npsL-upO5Ww}}} of the system typing \texttt{`HELLO WORLD'} before continuing to read.

The system performs inference on whole strings recursively.  Each string $s$ is associated to a user symbol $x$ (e.g. finger flexion, eye gaze movement or BCI response).  This association is referred to as the \textit{query} presented to the user.  The user produces the appropriate user symbol which the system estimates as $\hat{x}$.  This estimate is incorporated into the system via a Bayesian update.  If belief in a canidate string exceeds a threshold (e.g. $95\%$) then a decision is made.  Otherwise queries repeat until a canidate string achieves this threshold.

System performance depends on two query qualities.  First, queries should facilitate inference: they ought to yield evidence which disambiguates $s$ to the greatest extent possible.  Second, queries should be user-friendly: they ought to be easily understood by a non-technical user to allow quick, comfortable identification of the user symbol to be generated.  While our focus is constructing strong inference queries we put this objective aside for a moment.  To provide user-friendly queries we adopt a popular heuristic: strings are grouped via a prefix tree (i.e. a radix tree or trie).  An example over the alphabet `A', `B' and `.' is given below:
\begin{figure}[H]
	\centering
    \includegraphics[width=.25\textwidth]{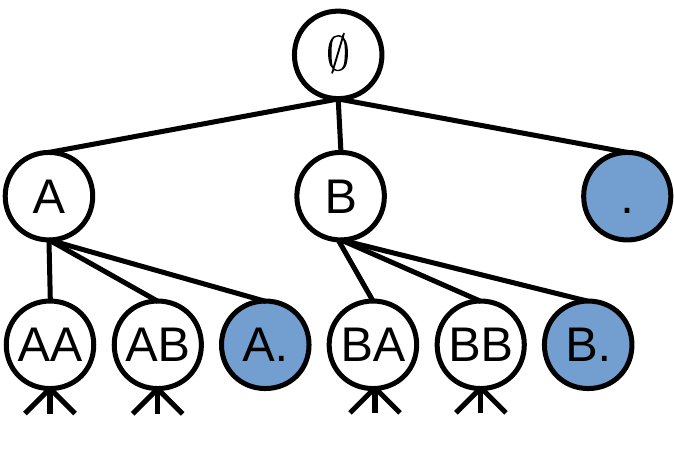}
    	\caption{An infinite prefix tree whose leafs are highlighted blue.\label{fig:str_tree2}}
\end{figure}
The name `prefix tree' derives from the fact that all descendants of a node are prefixed by its string.  Inspired by DASHER \cite{Wills2006} we use `.' as the end-of-message character which, by convention, concludes each $s$.  Because `.' must be the last character in a string it may only appear in a leaf, a node without children.  Note that the inclusion of this end-of-message character associates each string to exactly one leaf in the infinite prefix tree.

A query is defined by a mapping from all strings to user symbols, $S \rightarrow X$.  Of course, it isn't feasible to display the full infinite prefix tree.  Instead, the system selects a finite prefix tree.  The user is asked to identify the unique leaf in the finite tree which which prefixes their $s$.  Each leaf is associated to a (not-necesarily unique) user symbol.  Together, these steps yield $S \rightarrow X$ as required.  The system is summarized below:
\begin{figure}[H]
	\centering
    \includegraphics[width=.5\textwidth]{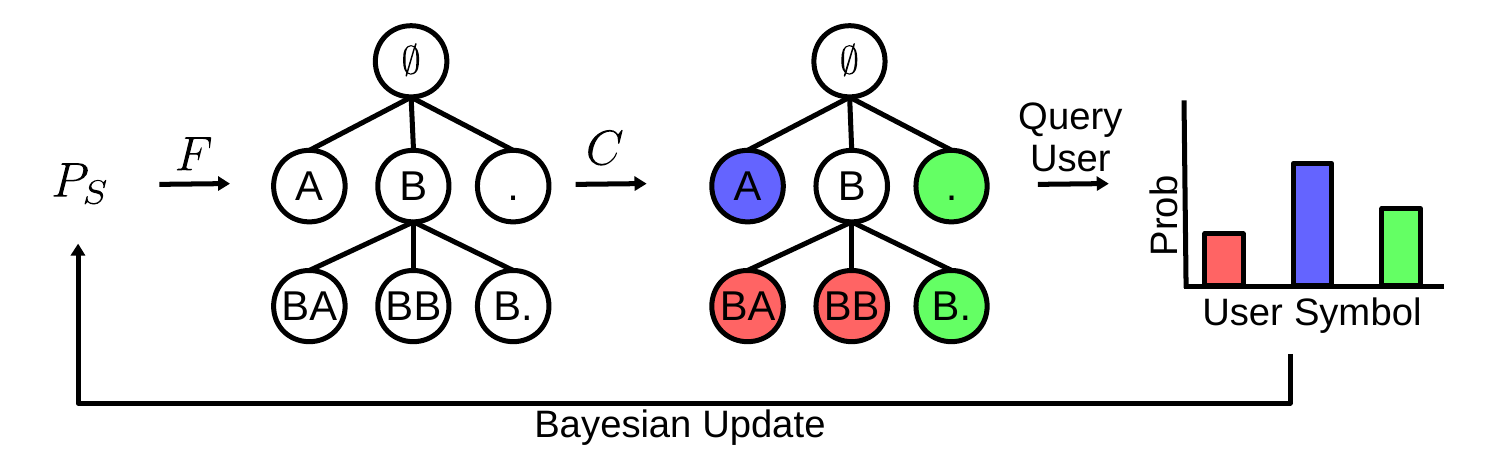}
        \caption{Query Loop.  If $s=\texttt{ABBA}$ then the user will be asked to generate the `green' user symbol.\label{fig:trial_flow}}
\end{figure}

More formally, a Bayes network describing this query structure is given in Fig \ref{fig:bayes_net}:
\begin{figure}[H]
  \centering
      \includegraphics[width=0.35\textwidth]{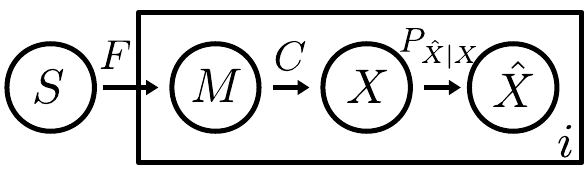}
  \caption{Inference model Bayes Net.  $i$ denotes trial index. \label{fig:bayes_net}}
\end{figure}
Selecting a finite prefix tree imposes a mapping $F$ from all strings to leafs of the finite prefix tree.  We use $M$ to denote these leafs.  Each leaf is associated with a user symbol via a coding function $C$.

\subsection{Objective Function}\label{ssec:obj}

As is typical \cite{cover2012elements, mackay_infoTheoryBook} of coding problems in Information Theory our objective is to achieve capacity by maximizing $I(S; \hat{X})$ where $I$ is the mutual information function.  It is helpful to note that:
\begin{align*} \label{eqn:obj}
I(S; \hat{X}) &= I(S; X; \hat{X}) + I(S; \hat{X} | X)
\\
&= I(S; X; \hat{X})
\\
&= I(S; X; \hat{X}) + I(X; \hat{X} | S)
\\
&= I(X; \hat{X}) \numberthis
\end{align*}
where the second equality relies on the fact that $I(S; \hat{X} | X) = 0$ since $S$ and $\hat{X}$ are independant given $X$ and the fourth uses the fact $I(X; \hat{X} | S) \leq H(X | S) = 0$ as there exists a deterministic function, $C \circ F$ which maps $S$ to $X$.  Please see Sec IV.D in \cite{higgerShuffle2016} for further intuitive motivation.  We add that this objective is identical (under a binary symmetric channel assumption) to the one used in Wills and MacKay's DASHER \cite{Wills2006} which makes use of arithmetic coding \cite{mackay_infoTheoryBook} for text entry.

\subsection{User Response Model ($P_{\hat{X}|X}$)}\label{ssec:user_response_model}
Because user symbol estimation is often noisy in practice it is helpful to incorporate evidence probabilistically.  Moreover, there is no guarantee that a user's symbols are equally accurate.  Furthermore, mistakes may not be distributed uniformly across other user symbols.  By characterizing $P_{\hat{X}|X}$ the system can encapsulate the varying accuracy and error structure of a user's inputs.  We estimate this distribution by normalizing a count of cross-validated training data for each $X$.  In other words we estimate $P_{\hat{X}|X}(x_i|x_j)$ as the percentage of the time which the system estimated an $x_j$ training sample as $x_i$.

Estimating this distribution serves two purposes.  First, it provides a principled way of incorporating user symbol evidence (see Sec \ref{ssec:bayes_update}).  Some user symbols are more reliably estimated than others and their detection should carry more weight than the detection of error prone user symbols.  Second, the characterization can be leveraged to optimize the query presented to the user (See Sec \ref{ssec:obj} and \ref{ssec:coding}).  This approach has demonstrated advantages over traditional decision tree style inference for single character selection \cite{higgerShuffle2016}.

\subsection{Bayesian update ($P_{S|\hat{X}}$)}\label{ssec:bayes_update}
To incorporate our observation of $\hat{x}$ into our belief about $S$ note that
\begin{equation}
\label{eqn:bayes_update}
P_{S|\hat{X}}(s|\hat{x}) = \frac{P_{\hat{X}|S}(\hat{x}|s)P_{S}(s)}{\sum_s P_{\hat{X}|S}(\hat{x}|s)P_{S}(s)}
\end{equation}
where
\begin{align*}
\label{eqn:likelihood}
P_{\hat{X}|S}(\hat{x}|s) &= \sum_m \sum_x P_{\hat{X}, X, M|S}(\hat{x}, x, m | s) \\
&= \sum_m \sum_x P_{\hat{X}|X}(\hat{x}|x) P_{X|M}(x|m) P_{M|S}(m|s) \\
&= \sum_m \sum_x P_{\hat{X}|X}(\hat{x}|x) \delta_{x, C(m)} \delta_{F(s), m} \\
&= P_{\hat{X}|X}(\hat{x}|C(F(s))). \numberthis
\end{align*}
The second equality uses the conditional independances shown in Fig \ref{fig:bayes_net}.  We use $\delta$ as the dirac delta function.  The third and fourth equalities make use of the fact that there exists a deterministic function $C \circ F: S \rightarrow X$.

\subsection{Selecting a Finite Prefix Tree}\label{ssec:prefix_tree_cut}
The selection of a finite prefix tree is equivilant to selecting a mapping $F$ from all strings to leafs of the finite prefix tree.  While choosing both $F$ and $C$ will affect our objective (\ref{eqn:obj}), we decouple the problem for computational ease.  Later on, we demonstrate that empirically (Sec \ref{sec:simulation}) and theoretically (\ref{sec:capacity}) this heuristic selection of $F$ does not preclude us from approaching the upper bound of our objective (\ref{eqn:obj}).

To offer a query which is as intuitive as possible to users, we cut the infinite prefix tree as:
\begin{equation}\label{eqn:fopt}
F_{opt} = \arg \max_F \min_m P_M(m)
\end{equation}
In other words, we cut the tree such that the leaf with minimum probability is as large as possible.  In doing so, we hope to offer a set of leafs which has the highest lower bound of `recognizability' from a user.

To approximate (\ref{eqn:fopt}), we begin with the empty node and \texttt{grow} and \texttt{prune} the tree as needed.  A \texttt{grow} adds every possible child of a leaf node.  A \texttt{prune} aggregates some set of sibling leafs into a representative node.  Examples are given below:

%Implementation-wise, probabilities are stored for every node in the tree (see blue text in Fig \ref{fig:split_group}).  Additionally, a conditional probability is maintained by \texttt{group} which remembers the relative weights among all the leafs represented in each aggregate node.  Defined as such these operations will preserve the probability of $S$ which includes information from the language model as well as all previous queries.

\begin{figure}[H]
  \centering
      \includegraphics[width=0.45\textwidth]{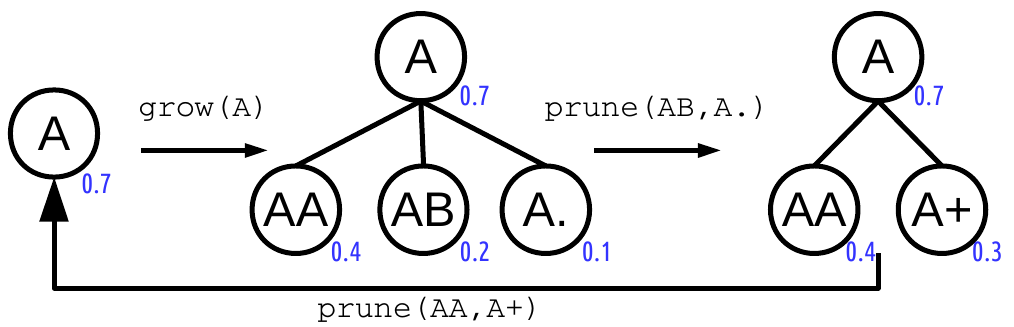}
  \caption{Example \texttt{grow} and \texttt{prune} operations over alphabet `AB.'.  An example $P_S$ is labelled in blue.  `+' is a wild card character which represents all siblings not shown.}
  \label{fig:split_group}
\end{figure}

Algorithm \ref{alg:optimizeF} greedily approximates $F_{opt}$ given in (\ref{eqn:fopt}), it is initialized with a graph consisting only of the empty node.  It accepts a user selected parameter, $L$, the maximum number of leafs present in the final finite prefix tree.
\begin{algorithm}
  \caption{Prefix Tree Selection \label{alg:optimizeF}}
  \begin{algorithmic}[1]
      \While{\textbf{exists} \texttt{leaf} \textbf{with} $P_S($\texttt{leaf}$) > \frac{1}{L}$}
          \State{\texttt{grow(leaf)}}
      \EndWhile
      \While{\texttt{num\_leaf} $> L$}
          \State{$s_1 \gets \arg \min_{s} P_S(s)$} \Comment{lowest prob}
          \State{$s_2 \gets \arg \min_{sib(s_1)} P_S(s)$} \Comment{sib of lowest prob}
          \State{\texttt{prune($s_1, s_2$)}}
      \EndWhile
  \end{algorithmic}
\end{algorithm} 

As is seen below, the number of leafs available to a tree effects system performance.  More leafs allows greater flexibility to optimize our objective (\ref{eqn:obj}) offering stronger inference.  On the other hand, too many leafs will stress a user's ability to find their target quickly (see the Hick-Hyman law in \cite{WilliamSoukoreff1995}).  We vary this parameter in simulation to quantify the inference benefit in increasing the number of leafs.

\subsection{Coding Function ($C: M \rightarrow X$)}\label{ssec:coding}
Given a finite prefix tree we seek to associate to each of its leafs a user symbol.  Remember that not all user symbol may offer equally reliable evidence.  For example, let us assume that $P_{\hat{X}|X}(x_1|x_1) >> P_{\hat{X}|X}(x_j|x_j)$ for all other $x_j \neq x_1$.  In other words, we can much more reliably detect when the user generates $x_1$ as compared to other user symbols.  In this case we priveldge a leaf by encoding it as $x_1$: if the user produces (or doesn't) $x_1$ we have strong evidence that their target is (or is not) a descendant of the leaf.  By assigning multiple leafs to $x_1$ we can extend this benefit though a new cost arises: received evidence is incapable from distinguishing among leafs assigned to the same user symbol.  We are fortunate that these consideration are encapsulated in the objective (\ref{eqn:obj}).

% introduce Q_x and greedy knapsack approach to coding

\subsection{Go-Back Node}
This section describe an aesthetic enhancement which is computationally equivilent to the method described above.  

Let us define the probability of a node as the belief that the node's string prefixes the user's intended message.  For leafs, this is simply $P_S(s)$ for the corresponding $s$.  Non-leaf nodes, those with children, are assigned probability as the sum of their children's probabilities.

Let us examine the probabilities along the path from the empty string node (i.e. the unique root node which has no parents) down to a user's target leaf $s$.  As defined above, this path must be monotonically decreasing in probability for every $P_S$.  After enough queries, the first few nodes in this path are shown to be promising prefixes while alternate paths are almost surely incorrect.  Visualization of all these dubious alternative paths is beastly and serves only to distract the user.

To clip away these low probability nodes from the tree we select a non-empty root node to begin our finite prefix tree.  In particular, we select the root as the longest string whose current belief exceeds some threshold.  By setting this threshold at $95\%$ we are garaunteed to avoid displaying strings which are not at least $5\%$ likely to prefix the target.  Of course, by doing this our new finite prefix tree no longer prefixes all possible strings.  To remedy this, we introduce a `go-back' node which stands in for all strings not represented in the tree.  The go-back node is so named because its selection may shorten the root node in future queries.   Evidence in favor of the go-back node is included probabilistically, its selection may not simply remove the final character of the root node string.  

\section{Web Speller}\label{sec:web_speller}
\begin{figure}
  \centering
      \includegraphics[width=0.4\textwidth]{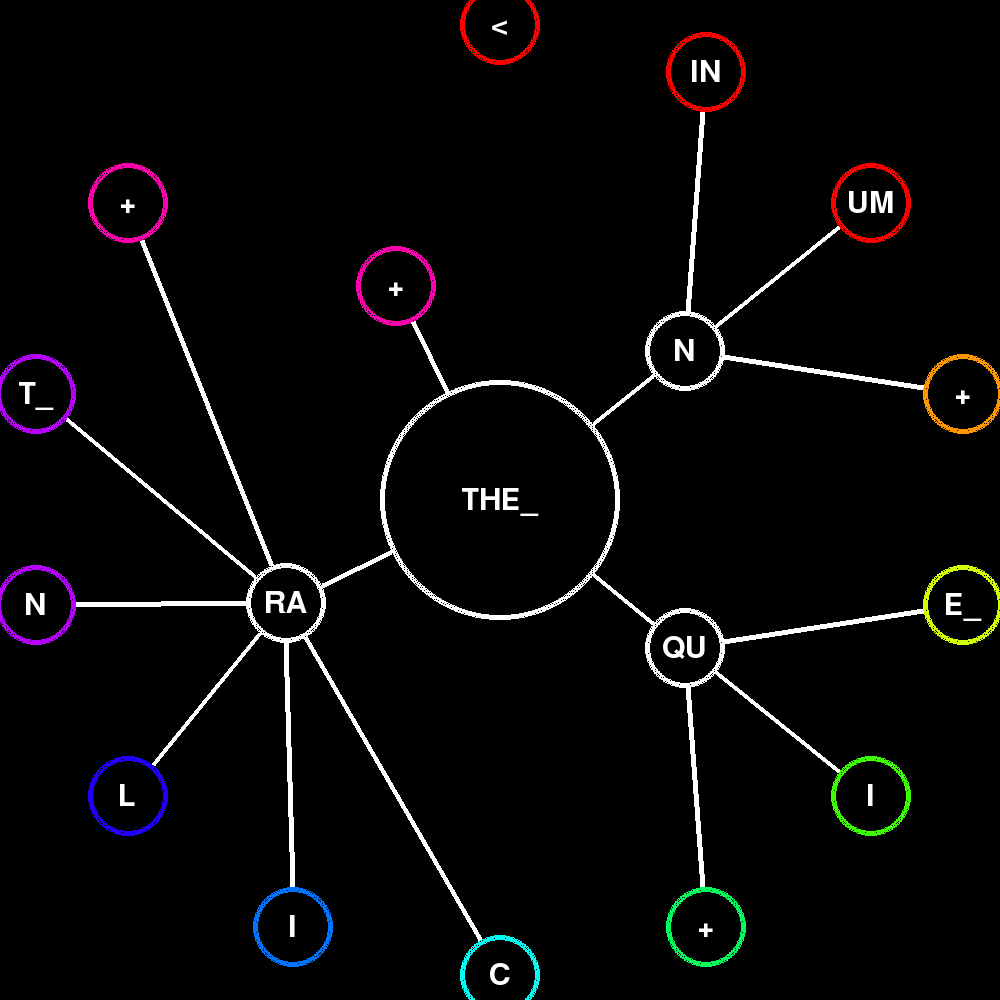}
  \caption{Screenshot of Web Speller interface.  A user who wishes to type `the quick brown fox' would target the green 'i' in the lower right; it represents all strings which are prefixed by 'the qui'.  \label{fig:web_speller_disp}}
\end{figure}
Users may have significant challenges identifying the leaf associated with their target string.  We describe the `Web Speller' which leverages angle inputs to perform string inference.  Such inputs are available via either traditional mouse, eye gaze or imagined movements.  It is straightforward to apply a similar decision scheme to other input modalities of interest to the Augmentative and Alternative Communication community (SSVEP, Motor Imagery or any typical AAC switch).

The Web Speller display (see video \footnote{\href{https://www.youtube.com/watch?v=r0P_5shu6GY}{\underline{https://www.youtube.com/watch?v=r0P\_5shu6GY}}})  positions the nodes such that the root node is in the center of the screen.  We position nodes such that radial distance to the root increases with tree depth and strings read clockwise are in alphabetical order.  Most significantly, an animation generates the next query tree from the current query tree.  For example, if a user's last leaf target was `A' then all nodes which are descendants of the `A', including their new target leaf, are animated from the previous leaf.  By doing so a user can expect their new target to appear at the location of their previous target.  We hope this design allows users to focus their attention on the appropriate subset of the animation to reduce the cognitive demands of the system.

Small (or imagined) movements are often available to a user so that they may produce angular directions of motion (\cite{Jarosiewicz2014}).  To make the Web Speller as intuitive as possible for these users, we position the leafs so that they are close to the angle of the associated user symbol.  In other words, if the user's target leaf is associated with a rightward angular input we position the leaf on the right side of the display.  Doing so imposes a second angular ordering in addition to the alphabetical clockwise constraint above.  We refer to a $C$ which meets this constraint as monotonic; in particular:
\begin{equation}
m_1 < m_2 \implies C(m_1) < C(m_2)
\end{equation}
where $<$ is defined as alphabetic ordering over $M$ and $C(m_i)$ are the user symbol angles in $[0, 2 \pi)$.  We examine the impact of this constraint on our objective (\ref{eqn:obj}) in Sec \ref{sec:simulation}.

\section{Simulation}\label{sec:simulation}
We test the efficacy of our multi-character approach against single-character querying by simulating a user typing a target sentence.  The user is assumed to have $10$ user symbols with:
\begin{equation}
\label{eqn:sim_conf_matrix}
P_{\hat{X}|X}(\hat{x}|x) = \left\{ \begin{matrix}
.9 \quad \text{if} \quad \hat{x} = x
\\
.011 \quad \text{otherwise}
\end{matrix}\right.
\end{equation}
Note that this input channel is maximized by a uniform $P_X$ and has capacity $2.54$ Bits.  (More complex accuracy and error structure is explored in \cite{higger_2016_tnsre}).  A user response is simulated by sampling from
\begin{equation}
\hat{X}_{sim} \sim P_{\hat{X}|X}(\hat{x}|C(F(s)))
\end{equation}

A 3-gram Witten-Bell \cite{Chen1996} language model is trained on the Brown Corpus, chosen for its availability and computational ease.  We reduce the corpus to only alphabetic characters, spaces and periods (the stopping character).  Model order is chosen to match the relatively modest size of the Brown Corpus.  This model offers $2.83$ Bits of entropy per character, computed by averaging over all contexts in the corpus.

We compare three modes.  The first is multi-character querying as described in Sec \ref{sec:method}.  The second is single-character querying which is identical to the multi-character scheme though Algorithm \ref{alg:optimizeF} is constrained to build prefix trees with depth 1.  (The system only queries about the next character).  The third mode imposes the monotonic constraint (see Sec \ref{sec:web_speller}) on the coding function $C$.  

The target string is `the quick brown fox jumps over the lazy dog'.  Within our model, this sentence is mildly more typical than average having $2.69$ Bits per character while the entire model had $2.83$.  We simulate while varying the numbers of leafs ($|M|$) from 3 to 16.  Each mode and $|M|$ attempts to type the target string $10$ times.  The string decision threshold is $95\%$.

\subsection{Results}\label{ssec:results}
\begin{figure}
  \centering
      \includegraphics[width=0.5\textwidth]{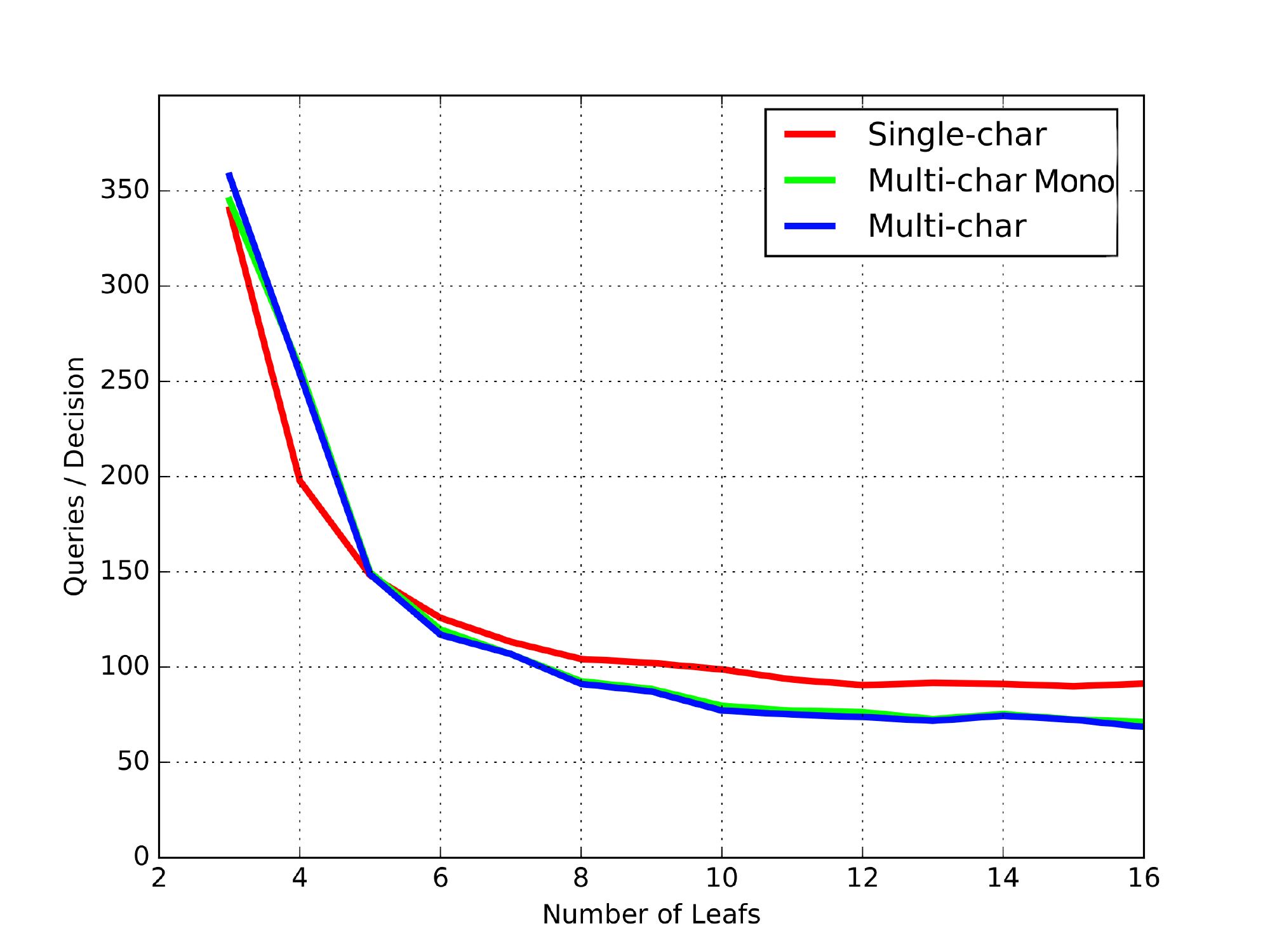}
  \caption{Number of leafs vs Queries per string decision.  Mutli-character querying is faster above some minimum number of leafs. \label{fig:speed}}
\end{figure}

Under all modes and $|M|$ the string is decided correctly.  This result makes sense given the high confidence needed to make a decision and the relatively low weight given to stopping characters in the language model.  As can be seen in Fig \ref{fig:speed} Multi-character querying uses $20.5\%$ fewer queries to produce the correct string decision compared to Single-character querying when $|M| \geq 10$.  This saturation point of $|M| = 10$ suggests there is little benefit to troubling the user by including more than 10 leafs on the display as there is marginal performance gained.

\begin{figure}
  \centering
      \includegraphics[width=0.5\textwidth]{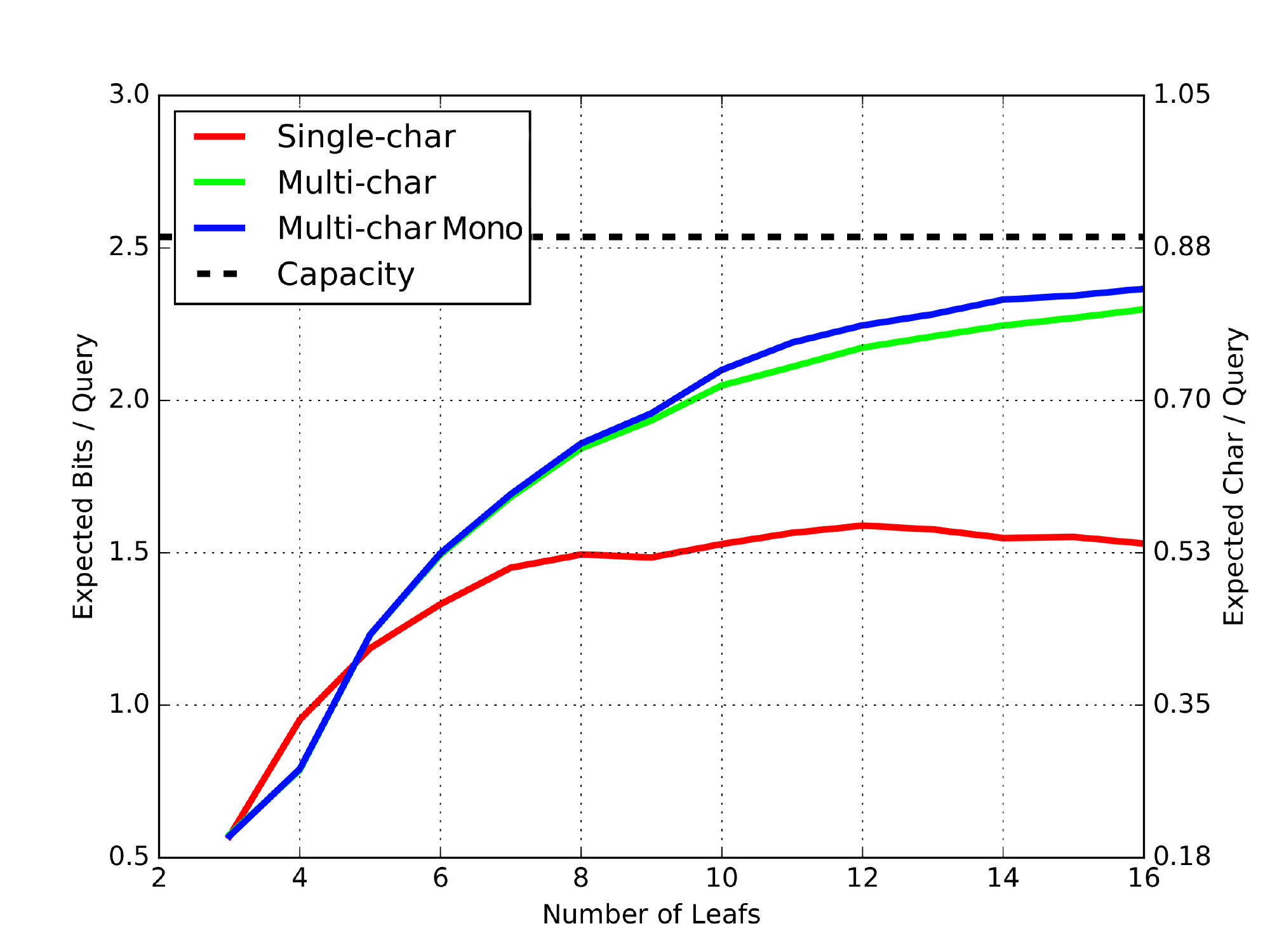}
  \caption{Number of leafs vs average expected Bits gained in a query.  To estimate the performance of our system on arbitrary text we construct a second y axis which assumes $2.85$ Bits of entropy per character (consistent with our language model).
 \label{fig:leafs_vs_rate}}
\end{figure}

Multi-character querying increases the expected entropy reduction (\ref{eqn:obj}) compared to single-character querying (Fig \ref{fig:leafs_vs_rate}).  Empirically, as the number of leafs increases both Multi-character query methods approach the capacity (see Sec \ref{sec:capacity}).

\begin{figure}
  \centering
      \includegraphics[width=0.5\textwidth]{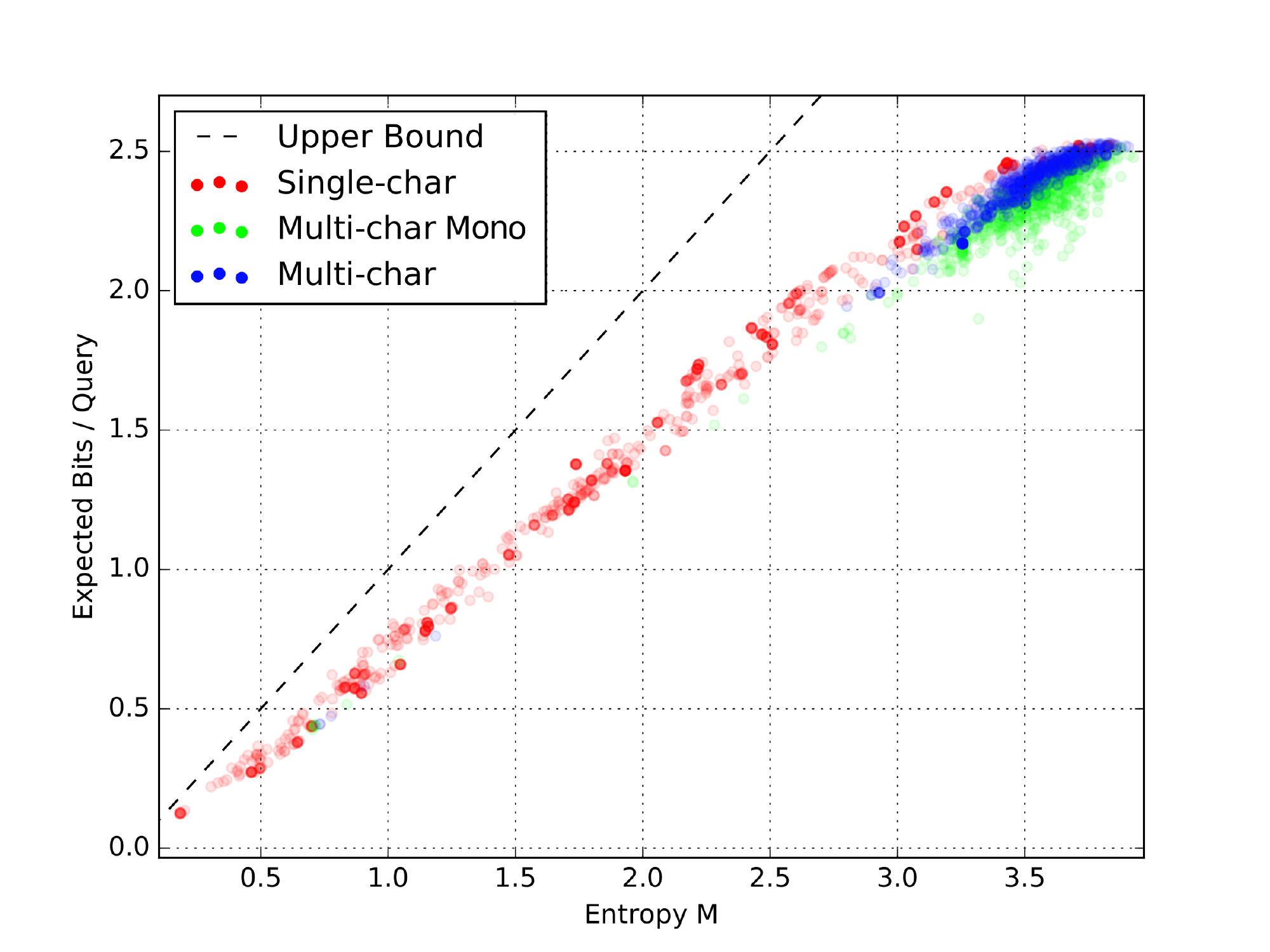}
  \caption{$H(M)$ vs expected bits gained in a query for all queries where $|M| = 16$.  Transparency is used to represent multiple queries which overlap (darker areas have more queries).  Motivation for upper bound given in (\ref{eqn:upper_bound}).  \label{fig:entM_vs_rate}}
\end{figure}

An upper bound for the expected Bits / Query is given by:
\begin{equation}
\label{eqn:upper_bound}
I(X; \hat{X}) = I(S; \hat{X}) \leq I(M; \hat{X}) \leq H(M)
\end{equation}
where the first equality is given by (\ref{eqn:obj}) and the first inequality follows from the Data Processing Inequality.  The system only ever learns what it asks about: queries about $M$ are necessarily limited to yield only the information available in $M$.  This phenomenon explains the relative weakness of single-character querying as seen in Fig \ref{fig:entM_vs_rate}.  In single-character querying the next character often has a low enough $H(M)$ to smother $I(S; \hat{X})$ but is not so low that the maximum probability among the next characters exceeds the decision threshold.  Note that this challenge also occurs near the end of Multi-character string decisions too, by definition the stopping character may not be \texttt{split} into smaller probabilities.  The few Multi-character queries with $H(M)$ less than 2 Bits in Fig \ref{fig:entM_vs_rate} immediately precede the final decision.

As can be seen in Fig \ref{fig:speed}, \ref{fig:leafs_vs_rate}, \ref{fig:entM_vs_rate} the monotonic constraint (see Sec \ref{sec:web_speller}) is fairly mild in limiting performance.  This suggests Web Speller can impose the monotonic constraint for greater visual clarity without too large an inference penalty.

\section{Achieving Capacity}\label{sec:capacity}
Assuming knowledge of a user's input model,  $P_{\hat{X}|X}$, the Web Speller algorithm reduces uncertainty in the distribution of intended user strings $s$ as quickly as possible (as $|M|$ diverges).  Let us define:
\begin{equation}
\label{eqn:opt_prior}
Q_x = \arg \max_{P_X} I(X; \hat{X}),
\end{equation}
the optimal prior distribution which achieves capacity $C = \max_{P_X} I(X; \hat{X})$.  Note that this distribution is globablly optimal without a constraint imposed to require that $Q_x$ be achievable by some $C$ and $F$ given $P_S$.  We can see from Algorithm \ref{alg:optimizeF} that:
\begin{equation}
\label{eqn:boundPm}
\max_m P_M(m) \leq \frac{2}{|M|}
\end{equation}
for all leafs except the go-back node (we delay inclusion of this detail until later).  This fact motivates our proof: if $P_M$ can be constructed to be arbitrarily fine grained (by letting $|M|$ diverge) then it may be pliable enough to create a $P_X$ arbitrarily close to $Q_X$ which yields the capacity.  We use $Q$ to denote the family of optimal distributions and $P$ to denote the family of achievable distributions. 

We claim that a greedy `water-filling' approach accomplishes this: map each $m$, in decreasing probability, to the $x$ whose current difference $Q_X(x) - P_X(x)$ is the greatest.  From this construction and (\ref{eqn:boundPm}) it can be seen:
\begin{equation}
\label{eqn:boundPQdiff}
\sum_x |Q_X(x) - P_X(x)| \leq \frac{2 |X|}{|M|}
\end{equation}
So that the two distributions converge as $|M|$ approaches infinity.  Note that this requires $Q_{\hat{X}}$ and $P_{\hat{X}}$, the distributions each imposes on $\hat{X}$, to also converge as the mapping between distributions on $x$ and $\hat{x}$ is continuous.  Similarly,  $H(Q_{\hat{X}})$ and $H(P_{\hat{X}})$ converge as the entropy function is also continuous.  Finally, we compute:
\begin{align*}
\label{eqn:proof_compute}
0 &= \lim_{|M| \rightarrow \infty} H(P_{\hat{X}}) - H(Q_{\hat{X}}) \\
&= \lim_{|M| \rightarrow \infty} E[\log Q(\hat{x})- \log P(\hat{x})] \\
&= \lim_{|M| \rightarrow \infty} E[\log \frac{P(\hat{x}|x)}{P(\hat{x})} - \log \frac{Q(\hat{x}|x)}{Q(\hat{x})}] \\
&= \lim_{|M| \rightarrow \infty} I_P(X; \hat{X}) - C \numberthis
\end{align*}
where we use the fact that $Q_{\hat{X}|X} = P_{\hat{X}|X}$  in the third equality and use $I_P(\cdot; \cdot)$ to represents the mutual information function under the distribution $P$.  This gives the desired result.

Note that the go-back node which may not be \texttt{split()} via algorithm \ref{alg:optimizeF}.  In this case (\ref{eqn:boundPm}) may not be valid.  If we add the following condition:
\begin{equation}
\max_m Q(m) < P_M(\text{`$<$'})
\end{equation}
then the above proof is still valid.  Note that this condition is relatively mild as:
\begin{equation}
P_M(\text{`$<$'}) \leq 1 - \alpha
\end{equation}
where `$<$' represents the go-back leaf and $\alpha$ is the decision threshold.

\section{Conclusion}\label{sec:conclusion}
We extend a single-character inference scheme which adapts to structure in the user errors as well as the language model to perform inference on the entire string.  Single-Character querying suffers from the fact that the information in the next undecided character may not satiate the Capacity of the user's input.  Our multi-character querying scheme adds in information from future characters to ensure that the system learns as much as possible with each query.  We prove this multi-character querying scheme approaches the upper bound (Capacity) of the user's input channel as queries are allowed to have more leafs.  However, as Hick-Hyman (\cite{WilliamSoukoreff1995}) informs us, providing the user with more options requires a longer time for the user to find their target.  Such considerations are paramount in developing the Web Speller into an interface which offers value to real users.

\section*{Acknowledgment}
Thanks to Betts Peters of Oregon Health and Science University.  If we offer any value to real users, it is because she helped us identify the problems worth solving.
 
\bibliographystyle{IEEEtran}
\bibliography{IEEEabrv,refs}

\end{document}